\documentstyle[12pt,epsfig]{article}
\begin{document}

\vspace*{0.5cm}

\begin{center}
{\large\bf The Higgs Boson: Shall We See It Soon Or Is It Still Far Away?} \\[1cm]%

{\bf D. I.~Kazakov } \\[5mm]

{\it BLTP, JINR, Dubna and ITEP, Moscow
 \\ e-mail: kazakovd@thsun1.jinr.ru}

\end{center}
\vspace{0.5cm}

\begin{flushright}
"The search for the Higgs boson is\\ the task \# 1 of high energy physics"\\
L.B.Okun', Talk at L.D.Landau Memorial Seminar,\\ Moscow, January 1998
~\cite{LB}
\end{flushright}
\vspace{1cm}

\begin{abstract} The status of the Higgs boson mass in the Standard Model
and its supersymmetric extensions is reviewed and the perspectives of Higgs
searches are discussed. The parameter space of the Minimal Supersymmetric
Standard Model (MSSM) is analysed with the emphasis on the  lightest Higgs
mass. The infrared behaviour of renormalization group equations for the
parameters of MSSM is examined and infrared quasi-fixed points are used for
the Higgs mass predictions.  They strongly suggest  the Higgs mass to be
lighter than 100 or 130 GeV for low and high $\tan\beta$ scenarios,
respectively. Extended models, however,  allow one to increase these limits
for low $\tan\beta$ up to 50\%.
\end{abstract}

\section{Introduction}

The last unobserved particle from the Standard Model is the Higgs boson. Its
discovery would allow one to complete the SM paradigm and confirm the
mechanism of spontaneous symmetry breaking. On the contrary, the absence of
the Higgs boson would awake doubts about the whole picture and would require
new concepts.

The Higgs mechanism is the simplest and minimal mechanism which allows one
to provide masses to all the particles of the SM, preserving the
renormalizability of a theory. It introduces a single new particle - the
Higgs boson, which is considered to be a point-like particle, or a bound
state in some approaches, and is supposed to be neutral and massive with the
mass of an order of the electroweak breaking scale, i.e. $10^2$ GeV.

Experimental limits on the Higgs boson mass come from a direct search at LEP
II and Tevatron and from indirect fits of
 electroweak precision data, first of all from the radiative
corrections to the W and top quark masses. A combined fit of modern
experimental data gives~\cite{EWWG}
\begin{equation}
 m_h=78^{+86}_{-47}\; {\rm GeV},
\end{equation}
 which at the 95\% confidence level leads to
the upper bound of 260 GeV. At the same time, recent direct
searches at LEP II for the c.m. energy of 189 GeV give the lower
limit of almost 95 GeV\cite{EWWG}.

Within the Standard Model the value of the Higgs mass $m_h$ is not
predicted. The effective potential of the Higgs field at the  tree level is
\begin{equation}\label{H}
  V_{eff}=-m^2|H|^2+\frac{\lambda}{2}(|H|^2)^2.
\end{equation}
The minimum of $V_{eff}$ is achieved for non-vanishing v.e.v. of the Higgs
field $<H>=v$ equal to $v=m/\sqrt{\lambda}$, which gives the mass $m_h =
\sqrt{2 \lambda} v$ as  a function of the vacuum expectation value of the
Higgs field, $v$ = 174.1 GeV, and the quartic coupling $\lambda$ which is a
free parameter.  However, one can get the bounds on the Higgs mass. They
follow from the behaviour of the quartic coupling which obeys the following
renormalization group equation describing the change of $\lambda$ with a
scale:
\begin{equation}\label{betalambda}
   \frac{d \lambda}{d t} = \frac{1}{16 \pi^2} \left(
6\lambda^2  + 6\lambda h_t^2 - 6h_t^4  + gauge \ terms \right)
\end{equation}
with $t= \ln (Q^2/\mu^2)$. Here $h_t$ is the top-quark Yukawa
coupling. Since the quartic coupling grows with rising energy
indefinitely, an upper bound on $m_h$ follows from the requirement
that the theory be valid up to the scale $M_{Planck}$ or up to a
given cut-off scale $\Lambda$ below $M_{Planck}$ \cite{R3}. The
scale $\Lambda$ could be identified with the scale at which a
Landau pole develops.  The upper bound on $m_h$ depends mildly on
the top-quark mass through the impact of the top-quark Yukawa
coupling on the running of the quartic coupling $\lambda$.

On the other hand, the requirement of vacuum stability in the SM (positivity
of $\lambda$) imposes a lower bound on the Higgs boson mass, which crucially
 depends on the top-quark mass as well as on the cut-off $\Lambda$ \cite{R3,R6}.
Again, the dependence of this lower bound on $m_t$ is due to the effect of
the top-quark Yukawa coupling on the quartic coupling in
eq.(\ref{betalambda}), which drives $\lambda$ to negative values at large
scales, thus destabilizing the standard electroweak vacuum.

From the point of view of LEP and Tevatron physics, the upper bound on the
SM Higgs boson mass does not pose any relevant restriction. The lower bound
on $m_h$, instead, is particularly important in view of search for the Higgs
boson at LEPII and Tevatron. For $m_t\sim  174$ GeV and
$\alpha_s(M_Z)=0.118$ the results at $\Lambda=10^{19}$ GeV or at $\Lambda=1$
TeV can be given by the approximate formulae\cite{R6}
\begin{eqnarray}
m_h&>&135
+2.1[m_t-174]-4.5\left[\frac{\alpha_s(M_Z)-0.118}{0.006}\right], \
\ \ \ \Lambda=10^{19} \  GeV, \label{19G}\\ m_h&>&72
+0.9[m_t-174]-1.0\left[\frac{\alpha_s(M_Z)-0.118}{0.006}\right], \
\ \ \ \Lambda=1 \ TeV, \label{1T}
\end{eqnarray}
where the masses are in units of  GeV.

\begin{figure}[htb]
\begin{center}
\epsfig{file=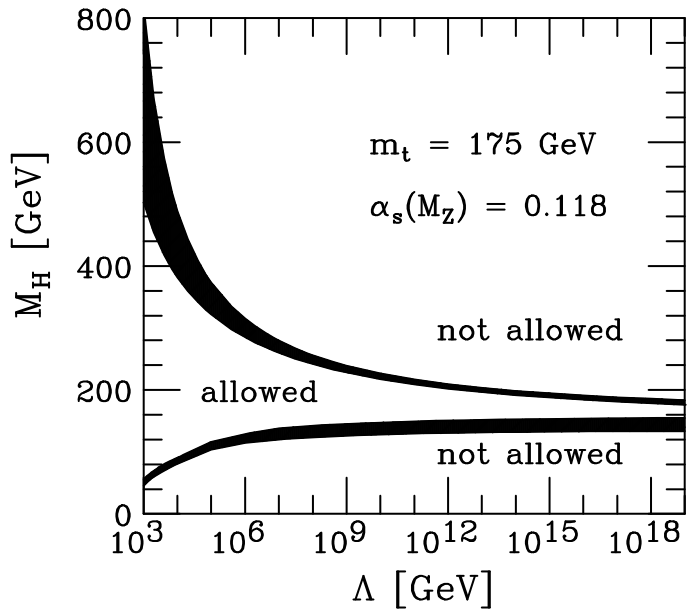,width=8cm,angle=90}
\end{center}
\caption{ Strong interaction and stability bounds on the SM Higgs
  boson mass. $\Lambda$ denotes the energy scale up to which the SM is valid.}
\label{mariano}
\end{figure}

Fig.\ref{mariano}~\cite{HR} shows the perturbativity and stability
bounds on the Higgs boson mass of the SM for different values of
the cut-off $\Lambda$ at which new physics is expected.
%
%
We see from Fig.\ref{mariano} and eqs.(\ref{19G},\ref{1T}) that
indeed for $m_t\sim 174$ GeV the discovery of a Higgs particle at
LEPII would imply that the Standard Model breaks down at a scale
$\Lambda$ well below $M_{GUT}$ or $M_{Planck}$, smaller for
lighter Higgs.  Actually, if the SM is valid up to $\Lambda \sim
M_{GUT}$ or $M_{Planck}$, for $m_t\sim 174$ GeV only a small range
of values is allowed: $134<m_h<\sim 200$ GeV.  For $m_t$ = 174 GeV
and $m_h < 100$~GeV [i.e. in the LEPII range] new physics should
appear below the scale $\Lambda \sim$ a few to 100~TeV. The
dependence on the top-quark mass however is noticeable. A lower
value, $m_t \simeq$ 170 GeV, would relax the previous requirement
to $\Lambda \sim 10^3 $ TeV, while a heavier value $ m_t \simeq$
180 GeV would demand new physics at an energy scale as low as
10~TeV.

The previous bounds on the scale at which new physics should
appear can be relaxed if the possibility of a metastable vacuum is
taken into account \cite{met}.   In this case, the lower bounds on
$m_h$ follow from requiring that no transition at any finite
temperature occurs, so that all space remains in the metastable
electroweak vacuum. In practice, if the metastability arguments
are taken into account, the lower bounds on $m_h$ become gradually
weaker, though the calculations become less reliable.

 On the other hand, this low limit is only valid in the SM
with one Higgs doublet:  it is enough to add a second doublet with the mass
lighter than $\Lambda$ to avoid it.  A particularly important example of a
theory where the bound is avoided is the Minimal Supersymmetric Standard
Model.

\section{The Higgs Boson Mass in Minimal Supersymmetry}

Supersymmetric extensions of the Standard Model  are believed to be the most
promising theories at high energies. An attractive feature of SUSY theories
is a possibility of unifying various forces of Nature.  The best known
supersymmetric extension of the SM is the   Minimal Supersymmetric Standard
Model (MSSM) \cite{1}. The parameter freedom of the MSSM comes mainly from
the so-called soft SUSY breaking terms, which are the sources of uncertainty
in the MSSM predictions. The most common way of reducing this uncertainty is
to assume universality of  soft terms, which means an equality of some
parameters at a high energy scale. Adopting the universality, one reduces
the parameter space to a five-dimensional one \cite{1}: $m_0$,  $m_{1/2}$,
$A$, $\mu$, and $B$. The last two parameters are convenient to trade for the
electroweak scale $v^2=v_{1}^2+v_{2}^2$, and $\tan\beta=v_{2}/v_{1}$, where
$v_{1}$ and $v_{2}$ are the Higgs field vacuum expectation values.

Contrary to the SM, in the MSSM there are at least two Higgs doublets. At
the tree level the Higgs potential containing the neutral components and
therefore responsible for the masses of physical scalars  has the form
\begin{equation}
V = m_1^2 |H_1|^2 + m_2^2 |H_2|^2 - m_3^2 \left(H_1H_2 +
h.c.\right) + \frac{g^2+g'^2}{8}\left(|H_1|^2 - |H_2|^2 \right)^2.
\end{equation}
 There are five physical eigenstates: $CP$-even Higgses $h$ and $H$,
 $CP$-odd Higgs $A$ and a pair of charged Higgses $H^\pm$, which at
tree level have the following masses
\begin{eqnarray}
m_A^2 &=& m_1^2 + m_2^2\ , \nonumber \\ m_{h,H}^2 &=& \frac12 \left[ m_A^2 +
M_Z^2 \mp \sqrt{\left( m_A^2 + M_Z^2 \right)^2 - 4 m_A^2 M_Z^2 \cos^2
2\beta} \right]\ , \\ m_{H^\pm}^2 &=& m_A^2 + M_W^2\ . \nonumber
\end{eqnarray}
For $m_A \gg M_Z$ the mass of the lightest Higgs is less than the $Z$ boson
mass
\begin{equation}\label{bound}
  m_h \simeq M_Z|\cos 2\beta|,
\end{equation}
 independently of any other parameters.  However, the inequality $m_h<M_Z$
is violated by radiative corrections, and the lightest Higgs mass
can exceed a hundred GeV, but not very much~\cite{2,3}.

The detailed analysis of the MSSM parameter space can be performed by
minimization of a $\chi^2$ function. This analysis implies also that a
number of constraints on the parameters like the  gauge coupling
unification, $b-\tau$ unification, radiative electroweak symmetry breaking,
dark matter density, $b\to s\gamma$ decay rate, etc are imposed. Details of
this analysis can be found in Ref.~\cite{we,G}.

\begin{figure}[htb]
\vspace*{-.8cm}
\begin{center}
\epsfig{figure=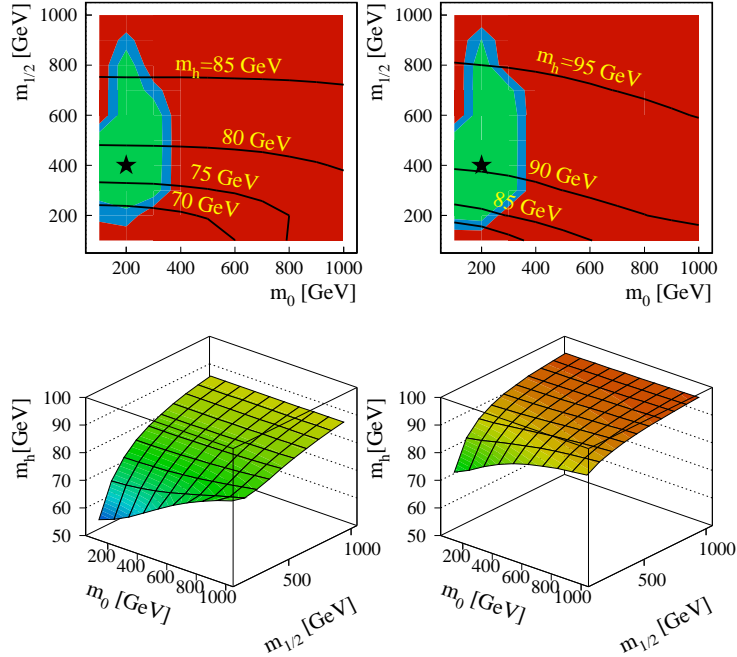, height=10cm}
\end{center}
\vspace{-1.0cm} \caption{Contours of the Higgs mass (solid lines) in the
$m_0,m_{1/2}$ plane (above) and the Higgs masses (below) for both signs of
$\mu$ for the low $\tan\beta$ solution $\tan\beta=1.65$ for $m_t=175$ GeV.}
\label{fig:2}
\end{figure}

For low $\tan\beta$ the present Higgs limit severely constrains the
parameter space, as can be seen from Fig.~\ref{fig:2}, which shows the
excluded regions in the ($m_0,m_{1/2})$ plane for different signs of $\mu$.
The experimental Higgs limit of 90 GeV is valid for the low $\tan\beta$
scenario ($\tan\beta<4$) of the MSSM too. As is apparent from
Fig.~\ref{fig:2} this limit clearly rules out the $\mu<0$ solution, in
agreement with other studies~\cite{abel}. However, this figure assumes
$m_t=175$ GeV. The top dependence on the Higgs mass is slightly steeper than
linear in this range and may move the contours within $\pm5$ GeV.

Adding about one $\sigma$ to the top mass, i.e. $m_t=180$ GeV, implies that
for the contours in Fig.~\ref{fig:2} one should add 6 GeV to the numbers
shown.  Even in this case the $\mu<0$ solution is excluded for a large
region of the parameter space. Only the small allowed region with
$m_{1/2}>700$ GeV is still available for $m_t=180$ GeV. Note that in this
region the squarks are well above  1 TeV, so in this case the cancellation
of the quadratic divergencies in the Higgs masses, which is only perfect if
sparticles and particles have the same masses, starts to become worrying
again.

For $\mu>0$ practically the whole plane  is allowed, except for the left
bottom corner shown on the top right-hand side of Fig.~\ref{fig:2}, although
the latest experiment has almost covered this region~\cite{EWWG}.

The  upper limit for the mass of the lightest Higgs is reached for heavy
squarks, but it saturates quickly, as is apparent from the bottom row in
Fig.~\ref{fig:2}.  For $m_0=1000,m_{1/2}=1000$, which corresponds to squarks
masses of about 2 TeV, one finds for the upper limit on the Higgs mass in
the low $\tan\beta$ scenario~\cite{pl}:
  $$ m_h^{max}=97\pm6~{\rm GeV}, $$
where the error is dominated by the uncertainty from the top mass.
 If one requires the squarks to be below 1 TeV, these upper limits are
reduced by 4 GeV.

For high $\tan\beta$ the upper limit on the Higgs mass in the Constrained
MSSM is~\cite{pl}
 $$ m_h^{max}=120\pm2~{\rm GeV}. $$
 The error from the top mass is small since the high $\tan\beta$ fits  anyway
prefer top masses around 190 GeV.

\section{Infrared Quasi-Fixed Point Scenario}

One of the possible ways to reduce the parameter freedom of the
MSSM is to use the fact that some low-energy  parameters are
insensitive to their initial high-energy values. This allows one
to find them without detailed knowledge of  physics at high
energies. To do this one has to examine the infrared behaviour of
renormalization group equations (RGEs) for these parameters and
use possible infrared fixed points to further restrict them.
Notice, however, that the true IR fixed points, discussed e.g. in
Ref.~\cite{PR} are reached only in the asymptotic regime. More
interesting is another possibility connected with so-called
infrared quasi-fixed points (IRQFPs) first discussed in
Ref.~\cite{6} and then widely studied by other
authors~\cite{7}-\cite{21}.  These fixed points usually give the
upper (or lower) bounds for the relevant solutions.

The well-known example of such infrared behaviour is the top-quark Yukawa
coupling $Y_t=h_t/(4\pi)^2$ for low $\tan\beta$ in the framework of the
MSSM. In this case the corresponding one-loop RGE has exact solution
\begin{equation}
Y_t(t)=\frac{Y_{0t}E(t)}{1+6Y_{0t}F(t)},
\end{equation}
where $E(t)$ and $F(t)$ are some known functions. It exhibits the
IRQFP behaviour in the limit $Y_{0t}=Y_t (0) \to
\infty$~\cite{6,9,10,11,13,14} where the solution becomes
independent of the initial conditions:
\begin{equation}
Y_t(t)\Rightarrow Y_t^{FP}=\frac{E(t)}{6 F(t)}. \label{fp} 
\end{equation}
A similar conclusion is valid for the other couplings
\cite{11,13,15,16,21,14}. It has been pointed out that the IRQFPs
exist for the trilinear SUSY breaking parameter $A_t$ \cite{13},
for the squark masses \cite{11,15} and for the other soft
supersymmetry breaking parameters in the Higgs and squark sector
\cite{21}.

In the case of large $\tan\beta$  the system of the RGEs has no
analytical solution and one can use either numerical or
approximate ones. It has been shown \cite{JK} that almost all SUSY
breaking parameters exhibit IRQFP behaviour.

For the IRQFP solutions the dependence on initial conditions
$Y_{i0},A_0$ and $m_0$ disappears at low energies. This allows one
to reduce the number of unknown parameters and make predictions
for the MSSM particle masses as functions the only free parameter,
namely $m_{1/2}$, or the gaugino mass, while the other parameters
are strongly restricted.

The strategy is the following~\cite{YJK,JK}. As input parameters
one takes the known values of the top-quark, bottom-quark and
$\tau$-lepton masses ($m_{t}, m_b, m_{\tau}$), the experimental
values of the gauge couplings \cite{EWWG} $\alpha_3=0.118,
\alpha_2=0.034, \alpha_1=0.017$, the sum of Higgs vev's squared
$v^2=v_1^2+v_2^2=(174.1\ GeV)^2$ and the  fixed-point values for
the
 Yukawa couplings and SUSY breaking parameters.
 To determine $\tan\beta$  the relations
between the running quark masses and the Higgs v.e.v.s in the MSSM are used
\begin{eqnarray} m_t &=& h_t \ v \ \sin\beta\,,
\label{mt} \\
m_b &=& h_b\  v \ \cos\beta\,, \label{mb} \\
m_{\tau} &=& h_{\tau}\  v \ \cos\beta\,. \label{mtau} 
\end{eqnarray}
The Higgs mixing parameter $\mu$ is defined from the minimization
conditions for the Higgs potential. Then, one is left with a
single free parameter, namely $m_{1/2}$, which is directly related
to the gluino mass $M_3$. Varying this parameter within the
experimentally allowed range, one gets all the masses  as
functions of this parameter.

For low $tan\beta$ the value of $\sin\beta$ is determined from
eq.(\ref{mt}), while for high $\tan\beta$ it is more convenient to
use the relation $\tan\beta =\frac{m_t}{m_b}\frac{h_b}{h_t}$,
since the ratio $h_t/h_b$ is almost a constant in the range of
possible values of $h_t$ and $h_b$.

For the evaluation of  $\tan\beta$ one first needs to determine
the running top- and bottom-quark masses. One can find them using
the well-known relations to the pole masses (see e.g.
\cite{SW,P,14}), including both QCD and SUSY corrections. For the
top-quark one has:
\begin{equation}
m_t(m_t)=\frac{m_t^{pole}}{1+ \left(\frac{\Delta m_t}{m_t}\right)_{QCD} +
\left(\frac{\Delta m_t}{m_t}\right)_{SUSY}}, \label{mtpole}
\end{equation}
where $ m_t^{pole}=(174.1 \pm 5.4)$ GeV  \cite{25}. Then,  the
following procedure is used to evaluate the running top mass.
First, only the QCD correction is taken into account and
$m_t(m_t)$ is found in the first approximation. This allows one to
determine both the stop masses and the stop mixing angle. Next,
having at hand the stop and gluino masses, one takes into account
the stop/gluino corrections.

For the bottom quark the situation is more complicated because the
mass of the bottom quark $m_b$ is essentially smaller than the
scale $M_Z$ and so one has to take into account the running of
this mass from the scale $m_b$ to the scale $M_Z$. The procedure
is the following \cite{P,26,27}: one starts with the bottom-quark
pole mass, $m_b^{pole}=4.94 \pm 0.15$ \cite{28} and finds the SM
bottom-quark mass at the scale $m_b$ using the two-loop $QCD$
corrections
\begin{equation}
m_b(m_b)^{SM}=\frac{m_b^{pole}}{1+ \left( \frac{\Delta m_b}{m_b}
\right)_{QCD}}\,.  \label{mbpole}
\end{equation}
 Then, evolving this mass to the scale $M_Z$  and using a numerical
solution of the two-loop  SM RGEs \cite{P,27} with
$\alpha_3(M_Z)=0.12$ one obtains $m_b(M_Z)_{SM}=2.91$ GeV. Using
this value one can  calculate the sbottom masses and then return
back to take into account the SUSY corrections from massive SUSY
particles
\begin{equation}
m_b(M_Z)= \frac{m_b(M_Z)^{SM}}{1+ \left( \frac{ \Delta m_b }{ m_b }
\right)_{SUSY} }\,.
 \label{mbsusy}
\end{equation}

When calculating the stop and sbottom masses one needs to know the Higgs
mixing parameter $\mu$.  For determination of this parameter one uses the
relation between the $Z$-boson mass and the low-energy values of $m_{H_1}^2$
and $m_{H_2}^2$ which comes from the minimization of the Higgs potential:
\begin{equation}
\frac{M_Z^2}{2}+\mu^2=\frac{m_{H_1}^2+\Sigma_1-
(m_{H_2}^2+\Sigma_2) \tan^2\beta}{\tan^2\beta-1}\,, \label{MZC}
\end{equation}
where $\Sigma_1$ and $\Sigma_2$ are the one-loop
corrections~\cite{G}. Large contributions to these functions come
from stops and sbottoms. This equation allows one to obtain the
absolute value of $\mu$, the sign of $\mu$ remains a free
parameter.

Whence the quark running masses and the $\mu$ parameter are found, one can
determine   the corresponding values of $\tan\beta$ with the help of
eqs.(\ref{mt},\ref{mb}). This gives in low and high $\tan\beta$ cases,
respectively
\begin{eqnarray*}
\tan{\beta}&=&1.47 \pm 0.15 \pm 0.05 \ \ \ for \ \mu>0\,,\\
 \tan{\beta}&=&1.56 \pm 0.15 \pm 0.05 \ \ \ for \ \mu<0\,,\\
\tan{\beta}&=&76.3 \pm 0.6 \pm 0.3 \ \ \ \ \ for \ \mu>0\,,\\
 \tan{\beta}&=&45.7 \pm 0.9 \pm 0.4 \ \ \ \ \ for \ \mu<0\,.
\end{eqnarray*}
The deviations  from the central value are connected with the experimental
uncertainties of the top-quark mass, $\alpha_3(M_Z)$ and uncertainty due to
the fixed point values of $h_t(M_Z)$ and $h_b(M_Z)$.

Having all the relevant parameters at hand it is possible to
estimate the masses of the Higgs bosons. With the fixed point
behaviour one has the only dependence left, namely on $m_{1/2}$ or
the gluino mass $M_3$. It is restricted only experimentally:
$M_3>144$ GeV \cite{EWWG} for arbitrary values of the squarks
masses.

Let us start with low $\tan\beta$ case. The masses of CP-odd, charged and
CP-even heavy Higgses increase almost linearly with $M_3$. The main
restriction comes from the experimental limit on the lightest Higgs boson
mass. It excludes $\mu <0$ case and for $\mu>0$ requires the heavy gluino
mass $M_3 \geq 750$ GeV. Subsequently one obtains
 $$m_A >844\ GeV, \ \ m_{H^\pm}>846 \ GeV, \ \ m_H
> 848\  GeV, \ \ \  for  \ \mu>0,$$
i.e. these particles are too heavy to be detected in the nearest
experiments.

For high $\tan\beta$ already the requirement of positivity of $m_A^2$
excludes the region with small $M_3$. In the most promising region  $M_3>1$
TeV ($m_{1/2}>300$ GeV) for the both cases $\mu>0$ and $\mu<0$ the masses of
CP-odd, charged and CP-even heavy Higgses are also too heavy to be detected
in the near future
 $$m_A>1100 \ \mbox{GeV  for} \ \
\mu>0, \ \ \ m_A>570 \ \mbox{GeV for}\ \ \mu<0,$$
 $$m_{H^{\pm}} > 1105 \
\mbox{GeV for} \ \  \mu>0, \ \ \   m_{H^{\pm}}> 575\ \mbox{GeV
for}\ \ \mu<0.$$
 $$m_H > 1100 \ \mbox{GeV for} \ \  \mu>0, \ \ \ m_H > 570 \
\mbox{GeV  for}
 \ \ \mu<0.$$

\begin{figure}[p]
\vspace{-8cm}
  \begin{center}
\epsfig{figure=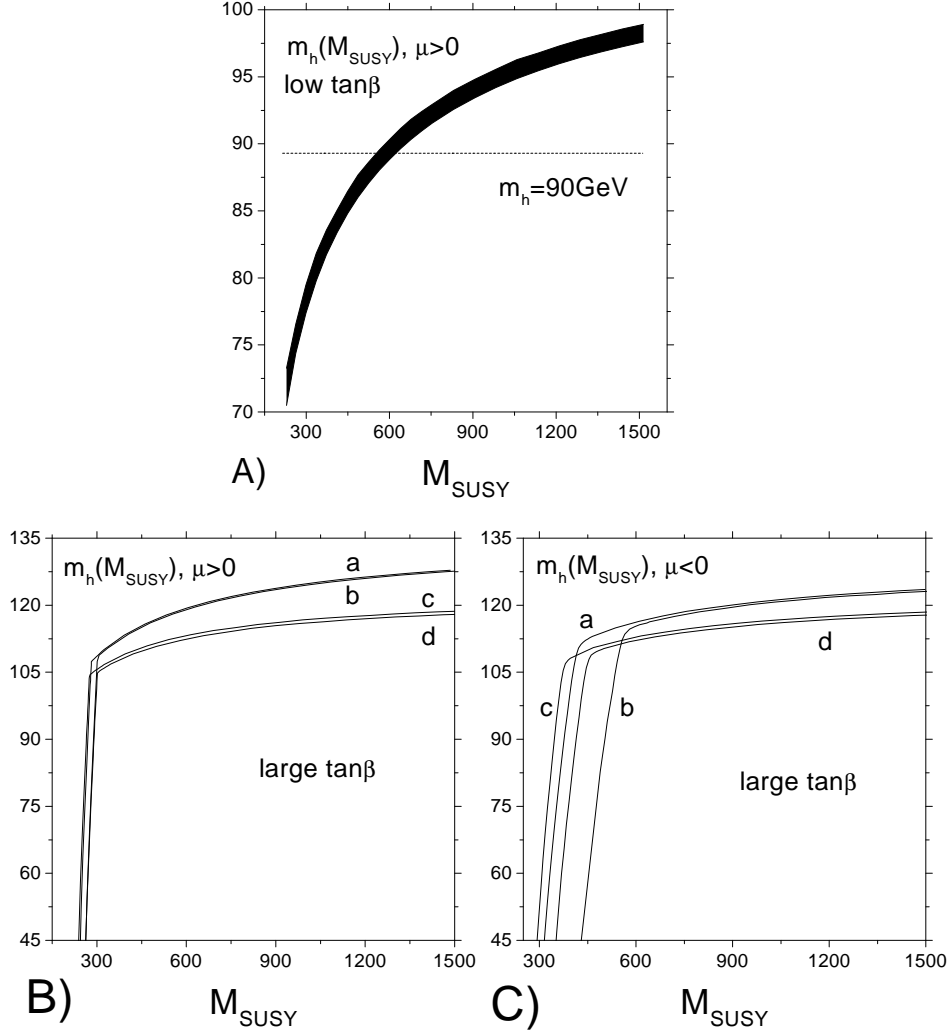, height=22cm}
  \end{center}
\caption{{\bf A)} The dependence of the mass of the lightest Higgs
boson $h$ on $M_{SUSY}=(\tilde m_{t_1} \tilde m_{t_2})^{1/2}$
(shaded area) for $\mu>0$, low $\tan\beta$.  The dashed line
corresponds to the minimum value of $m_h=90$ GeV allowed by
experiment.  {\bf B),C)} The mass of the lightest Higgs boson $h$
as function of $M_{SUSY}$ for different signs of $\mu$, large
$\tan\beta$. The curves (a,b) correspond to the upper limit of the
Yukawa couplings and to $m_0^2/m_{1/2}^2=0$ (a) or to
$m_0^2/m_{1/2}^2=2$ (b). The curves (c,d) correspond to the lower
limit of the Yukawa couplings and  to $m_0^2/m_{1/2}^2=0$ (c) or
 to $m_0^2/m_{1/2}^2=2$ (d).  Possible values of the mass of the
lightest Higgs boson are inside the areas marked by these lines.}
\label{fig:4}
\end{figure}

The situation is different for the lightest Higgs boson $h$, which
is much lighter. As has been already mentioned, for low
$\tan\beta$ the negative values of $\mu$ are excluded by the
experimental limits on the Higgs mass. Further on we consider only
the positive values of $\mu$. Fig.\,\ref{fig:4} shows the value of
$m_h$ for $\mu>0$ as a function of the geometrical mean of stop
masses - this parameter is often identified  with  a supersymmetry
breaking scale $M_{SUSY}$. One can see that the value of $m_h$
quickly saturates close to $\sim$ 100 GeV. For $M_{SUSY}$ of the
order of 1 TeV the value of the lightest Higgs mass is~\cite{YJK}
\begin{equation}
m_h=(94.3+1.6+0.6\pm5\pm0.4) \ \mbox{GeV}, \ \ \ \mbox{ for} \ M_{SUSY}=1 \
TeV, \label{mass}
\end{equation}
 where the first uncertainty comes from the deviations from the IRQFPs
 for the mass parameters, the second one is related to that of the
top-quark Yukawa coupling, the third reflects the uncertainty of the
top-quark mass of 5.4 GeV, and the last one comes from that of the strong
coupling.

One can see that the main source of uncertainty is the experimental error in
the top-quark mass. As for the uncertainties connected with the fixed
points, they give much smaller errors of the order of 1 GeV.

Note  that the  obtained result (\ref{mass}) is very close to the
upper boundary, $m_h=97$ GeV, obtained in Refs. \cite{14,pl} (see
the previous section).

For the high $\tan\beta$ case the lightest Higgs is slightly
heavier, but the difference is crucial for LEP II. The mass of the
lightest Higgs boson as a function of   $M_{SUSY}$ is shown in
Fig.\ref{fig:4}. One has  the following values of $m_h$ at a
typical scale $M_{SUSY}=1$ TeV ($M_3 \approx 1.3$TeV)~\cite{JK}:
\begin{eqnarray}
m_h&=&128.2 -0.4 - 7.1 \pm 5 \ \mbox{GeV, \ for} \  \mu>0 \,,
\nonumber \\ m_h&=&120.6 -0.1 - 3.8 \pm 5 \ \mbox{GeV,\ for} \
\mu<0 \,.  \nonumber
\end{eqnarray}
The first uncertainty is connected with the deviations from the IRQFPs for
mass parameters, the second one with the Yukawa coupling IRQFPs, and the
third one is due to the experimental uncertainty in the top-quark mass. One
can immediately see that the deviations from the IRQFPs for mass parameters
are negligible and only influence the steep fall of the function on the
left, which is related to the restriction on the CP-odd Higgs boson mass
$m_A$.  In contrast with the low $\tan\beta$ case, where the dependence on
the deviations from Yukawa fixed points was about $1$ GeV, in the present
case it is much stronger. The experimental uncertainty in the strong
coupling constant $\alpha_s$ is not included  because it is negligible
compared to those of the top-quark mass and the Yukawa couplings and is not
essential here contrary to the low $\tan\beta$ case.

One can see that for large $\tan\beta$ the masses of the lightest
Higgs boson are typically around 120 GeV that is too heavy for
observation at LEP II.

\section{Summary and Conclusion}

Thus, one can see that in the IRQFP approach all the Higgs bosons except for
the lightest one are found to be too heavy to be accessible in the nearest
experiments. This conclusion essentially coincides with the results of more
sophisticated analyses. The lightest neutral Higgs boson, on the contrary is
always light. In the case of low $\tan\beta$ its mass is small enough to be
detected or excluded  in the next two years when the c.m.energy of LEP II
reaches 200 GeV. On the other hand, for the high $\tan\beta$ scenario the
values of the lightest Higgs boson mass are typically around 120 GeV, which
is too heavy for the observation at LEP II leaving hopes for the Tevatron
and LHC.

\begin{figure}[htb]
\begin{center}\vspace*{-2cm}
\epsfig{file=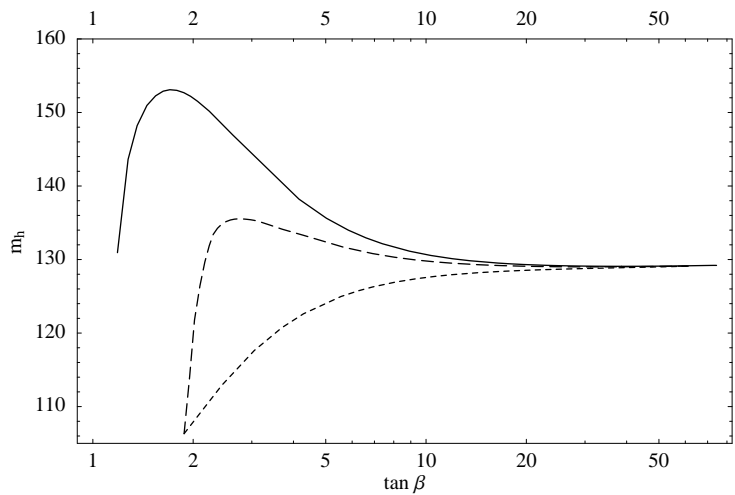,width=14cm}
\end{center}\vspace*{-10cm}
\caption{ Dependence of the upper bound on the lightest Higgs
boson mass on $\tan\beta$ in MSSM (lower curve), NMSSM (middle
curve) and extended SSM (upper curve)~\cite{pomarol}} \label{f3}
\end{figure}

However, these SUSY limits on the Higgs mass may not be so restricting if
non-minimal SUSY models are considered. In a SUSY model extended by a
singlet, the so-called Next-to-Minimal model, eq.(\ref{bound}) is modified
and at the tree level the bound looks like~\cite{pomarol}
\begin{equation}
  m_h^2 \simeq M_Z^2\cos^2 2\beta+ \lambda^2v^2\sin^2 2\beta,
\end{equation}
where $\lambda$ is an additional singlet Yukawa coupling. This
coupling being unknown brings us back to the SM situation, though
its influence is reduced by $\sin 2\beta$. As a result, for low
$\tan\beta$ the upper bound on the Higgs mass is slightly modified
(see Fig.\ref{f3}).

Even more dramatic changes are possible in models containing
non-standard fields at intermediate scales. These fields appear in
scenarios with gauge mediated supersymmetry breaking. In this
case, anyway the upper bound on the Higgs mass may increase up to
155 GeV~\cite{pomarol} (the upper curve in Fig.\ref{f3}), though
it is not necessarily saturated.  One should notice, however, that
these more sophisticated models do not change the generic feature
of SUSY theories, the presence of the light Higgs boson.

 \vglue 0.5cm

{\bf Acknowledgments}

\vglue 0.5cm

The author is grateful to A.V.Gladyshev  and M.Jur\v{c}i\v{s}in
for useful discussions and help in preparing the manuscript.
Financial support from RFBR grant \# 98-02-17453 is acknowledged.

\end{document}